# Dynamics of Symmetry-Breaking Stacking Boundaries in Bilayer MoS$_2$

Aiming Yan,[†,‡,§] Chin Shen Ong,[†,‡] Diana Y. Qiu,[†,‡] Colin Ophus,[∥] Jim Ciston,[∥] Christian Merino,[†] Steven G. Louie,[†,‡] and Alex Zettl*[,†,‡,§]

[†]Department of Physics, University of California, Berkeley, California 94720, United States
[‡]Materials Sciences Division, Lawrence Berkeley National Laboratory, Berkeley, California 94720, United States
[§]Kavli Energy NanoSciences Institute at the University of California, Berkeley and the Lawrence Berkeley National Laboratory, Berkeley, California 94720, United States
[∥]National Center for Electron Microscopy, Molecular Foundry, Lawrence Berkeley National Laboratory, Berkeley, California 94720, United States

**S** Supporting Information

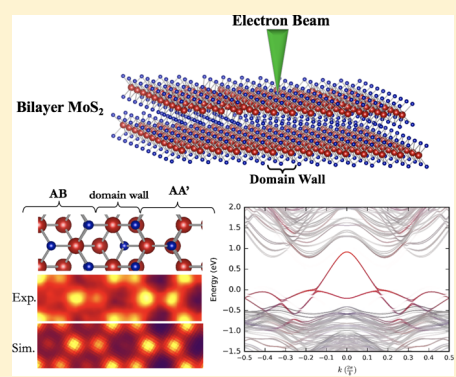

**ABSTRACT:** Crystal symmetry of two-dimensional (2D) materials plays an important role in their electronic and optical properties. Engineering symmetry in 2D materials has recently emerged as a promising way to achieve novel properties and functions. The noncentrosymmetric structure of monolayer transition metal dichalcogenides (TMDCs), such as molybdenum disulfide (MoS$_2$), has allowed for valley control via circularly polarized optical excitation. In bilayer TMDCs, inversion symmetry can be controlled by varying the stacking sequence, thus providing a pathway to engineer valley selectivity. Here, we report the in situ integration of AA′ and AB stacked bilayer MoS$_2$ with different inversion symmetries by creating atomically sharp stacking boundaries between the differently stacked domains, via thermal stimulation and electron irradiation, inside an atomic-resolution scanning transmission electron microscopy. The setup enables us to track the formation and atomic motion of the stacking boundaries in real time and with ultrahigh resolution which enables in-depth analysis on the atomic structure at the boundaries. In conjunction with density functional theory calculations, we establish the dynamics of the boundary nucleation and expansion and further identify metallic boundary states. Our approach provides a means to synthesize domain boundaries with intriguing transport properties and opens up a new avenue for controlling valleytronics in nanoscale domains via real-time patterning of domains with different symmetry properties.

Crystal symmetry in a material dictates its physical properties. Manipulating its symmetry provides a pathway to achieve novel and unusual functionalities. This is particularly relevant for two-dimensional (2D) materials, promising candidates for numerous applications, including next-generation flexible electronics. In bilayer graphene, for example, symmetry-inversion at a "soliton" stacking boundary between AB and BA stacked regions[1,2] leads to robust, topologically protected 1D conducting channels at the domain walls.[3−5] In other 2D materials such as transition metal dichalcogenides (TMDCs), the noncentrosymmetric crystal structure of the monolayer form leads to valley selectivity and holds promise for applications in spintronics and valleytronics.[6−9]

One prototypical example of 2D TMDCs is monolayer H-phase MoS$_2$. Monolayer H-phase MoS$_2$ is noncentrosymmetric and has a lattice constant of $a$ = 3.16 Å (Figure 1). The broken inversion symmetry, in addition to strong spin−orbit interactions, leads to the splitting of valence bands and spin-valley coupling,[6] enabling valley selectivity when excited by circularly polarized light.[7−9] However, inversion symmetry is restored in bilayer 2H-phase MoS$_2$ — a phase that is naturally found and most commonly observed (also known to adopt AA′ stacking sequence in Figure 1). To break the inversion symmetry in bilayer MoS$_2$, a vertical electrical field can be applied.[10] Another way to introduce asymmetry in bilayer and multilayer MoS$_2$ is to engineer the stacking sequence by folding[11] or transferring[12,13] exfoliated single-layer MoS$_2$, or by using chemical synthesis methods such as chemical vapor transport (CVT)[14] and chemical vapor deposition (CVD).[15−18] Two commonly observed stacking sequences in CVD/CVT grown MoS$_2$ are AA′ stacking and AB stacking (a representative schematic of AB stacking is shown in Figure 1), which are nearly energetically degenerate.[15,19] Repeated AB stacking forms the 3R phase of MoS$_2$, in which centrosymmetry is broken throughout the bulk. A recent study shows that valley-dependent spin polarization is realized in CVT synthesized 3R-phase bulk MoS$_2$.[14]

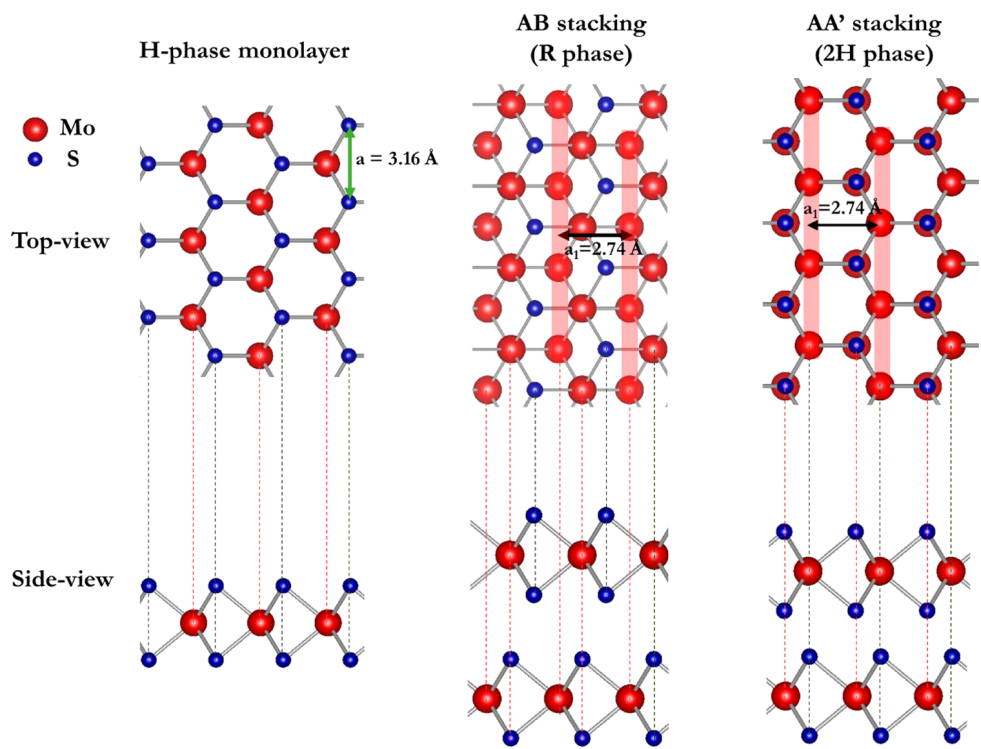

**Figure 1.** Ball-and-stick models for monolayer *H*-phase MoS$_2$, and AB (*R* phase) and AA′ (2*H* phase) stacked bilayer MoS$_2$, viewed from the top and the side, with red balls representing Mo atoms and blue balls representing S atoms. The lattice constant (*a*) for monolayer *H*-phase MoS$_2$ is 3.16 Å, and the Mo lattice distance in one layer MoS$_2$ ($a_1$) is 2.74 Å, as indicated by red shadings in the schematics for both AB and AA′ bilayers.

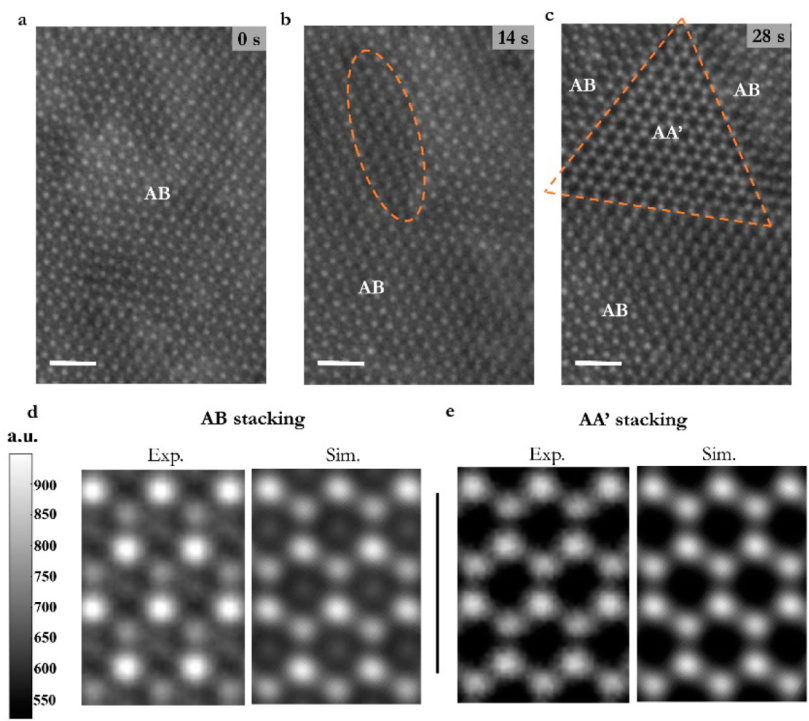

**Figure 2.** Stitching of AB and AA′ stacked regions with sharp stacking boundaries in bilayer MoS$_2$. (a) High-resolution STEM ADF image taken at time 0 s, showing AB stacked bilayer MoS$_2$. (b) After 14 s of scanning in the same area, rearrangement of atoms occurred, as highlighted by the orange ellipse at the center of this AB stacked region. (c) High-resolution STEM ADF image showing a triangular AA′ stacked region (highlighted by an orange triangle) surrounded by AB stacked region. This image was taken 14 s after (b) and is from the same area as (b). (d, e) STEM ADF images from AB stacked region in (a) and AA′ stacked region in (c), respectively, with the intensity averaged over more than 10 unit cells of the same stacking sequence. The experimental results are compared to the multislice simulation[26] of STEM ADF images of AB and AA′ stacked bilayer MoS$_2$ under the same experimental imaging condition in (d) and (e), respectively. All images in (d) and (e) are plotted on the same absolute scale. Scale bars in (a)−(c) are 1 nm.

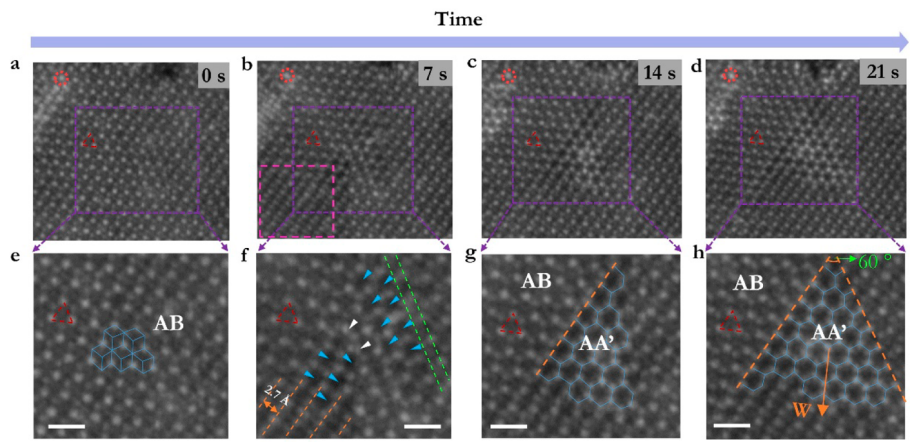

**Figure 3.** Nucleation and motion of the stacking boundaries in bilayer MoS$_2$ at 400 °C. (a)−(d) STEM ADF image series of structural change in bilayer MoS$_2$ from the same area as time evolves from left to right. Red circles and triangles in these images highlight the same atoms and lattice points in this area. (e)−(h) Enlarged images of the area highlighted with purple squares in (a)−(d), respectively. (e) At $t = 0$ s, AB stacked bilayer MoS$_2$ with the lattice points connected by blue lines. The S-S columns from the top layer occupy the hollow centers of the hexagonal lattices in this image. (f) At $t = 7$ s, local atomic rearrangement starts to occur, which triggers the nucleation of AA′ stacked domain. Large-area linear features with spacing 2.74 Å (orange lines) are due to the interlayer lateral shift between the top and bottom MoS$_2$ layers. Other structural changes are also observed, such as extra atoms (white arrowheads) and local shrinkage of projected Mo−Mo distance (green lines). (g) At $t = 14$ s, AA′ stacked region grows to ∼2.88 nm$^2$. (h) At $t = 21$ s, AA′ stacked region grows to 4.75 nm$^2$ as outlined by the blue hexagons. The angle between the two straight stacking boundaries is 60°. The growth direction for the AA′ stacked region is now along vector **W**. Scale bars in (e)−(h) are 5 Å.

In this work, we demonstrate a novel way in which the symmetry of bilayer MoS$_2$ can be engineered via the creation of nanoscale stacking boundaries that separate domains with different inversion symmetries. We show that, in bilayer MoS$_2$, the transition from AB to AA′ stacking or from AA′ to AB stacking can be realized by heating and electron-irradiating the AA′ or AB stacked bilayer MoS$_2$ inside an aberration-corrected scanning transmission electron microscope (STEM). By using *in situ* atomic-resolution STEM annular-dark field (ADF) imaging, we identify atomically sharp stacking boundaries between these two differently stacked domains. Combined with first-principles density functional theory (DFT) calculations, we discuss the atomic-scale dynamics of the domain nucleation and growth. DFT calculation also reveals the existence of highly localized metallic states at the domain boundaries.

Bilayer MoS$_2$ studied here is grown on SiO$_2$/Si substrates via a modified CVD method based on refs 20 and 21 (see the Supporting Information for details). The as-grown MoS$_2$ flakes are then transferred to a Protochip heating TEM grid via the PMMA transfer method.[22] The Protochip TEM grid holder has the ability to heat the sample up to 1000 °C with minimal local temperature variation. We use atomic-resolution STEM ADF imaging operated at 80 keV to track the stacking sequence change in bilayer MoS$_2$. STEM ADF images show the 2D projection of atomic positions of the layered material, and the intensity in STEM ADF images is closely related to the atomic number and thickness of the material. Since Mo and S have significantly different atomic numbers (Mo: 42 and S: 16), STEM ADF imaging is an ideal tool to identify the atomic species at different lattice positions, thus revealing the atomic structure of single- and few-layer MoS$_2$. High-resolution STEM ADF imaging has been used to visualize the atomic arrangement in single- and few-layer MoS$_2$, including stacking sequences, grain boundaries, point defects, edge sites, and even phase transformation.[15,20,21,23,24] In our study, the electron beam serves two functions: (1) it facilitates *in situ* imaging of local atomic structure and changes thereof in bilayer MoS$_2$; (2) it provides energy (in addition to the high temperature thermal bath) to stimulate local atomic movement. To make this atomic movement slow enough for real-time STEM tracking, an electron dose as low as $7 \times 10^6$ electrons/nm$^2$ s is applied. Although the electron energy of 80 keV is close to the threshold for S vacancy formation in monolayer MoS$_2$,[25] we are still able to image the regions of interest in bilayer samples for over 3 min. Movies composed of continuous STEM ADF imaging allows the transition between different stacking sequences to be recorded.

In our *in situ* STEM study, we start from a CVD-grown pristine MoS$_2$ bilayer with a uniform stacking sequence of AB or AA′. Transitions between different stacking sequences are observed at 350 and 400 °C. Below, we focus on the representative case where the AB stacking is locally transformed into AA′ stacking at 400 °C. Different stacking sequences are identified by comparing the experimental STEM ADF image with multislice simulation results, as shown in Figure 2d,e (see the Supporting Information for details). There are three distinct lattice points in the original AB stacked bilayer MoS$_2$ (Figure 1): (1) the highest intensity corresponds to the 2D projection of the Mo-S-S column (with Mo from the top layer and S-S from the bottom layer); (2) the second highest intensity corresponds to the Mo column from the bottom layer; and (3) the weakest intensity corresponds to the S-S column projected from the top layer. When viewed from the top (Figure 1), the Mo-S-S columns and Mo columns are configured in a hexagonal lattice with 3-fold symmetry, while the hollow centers of the hexagonal lattice are occupied by the S-S columns. As the electron beam scans in the AB stacked region (as shown in Figure 2a,d) for 14 s, local atomic rearrangement occurs (marked by an orange ellipse in Figure 2b). Strikingly, after another 14 s of scanning in the same area, a triangular region with AA′ stacking appears as shown in Figure 2c,e. In this case, the three stacking boundaries between AA′ and AB stacked regions form extremely rapidly and stay stationary as the scan continues.

In contrast, stacking boundaries can also be induced more gradually by using a lower electron beam intensity. Here, to study the dynamics of the transition between stacking sequences, continuous STEM ADF images are recorded as shown in Figure 3. Figure 3a–d shows the evolution of a fixed area in the bilayer $MoS_2$ sample in increments of 7 s. As a position reference between frames, red dotted circles mark the same atom from panel (a) to (d) in Figure 3, while dashed red triangles connect the same three lattice points from panel (a) to (h) in Figure 3. It is instructive to consider the nucleation of the stacking boundaries. Panels (a) and (e) in Figure 3 (Figure 3e is an enlarged image of the purple-squared area in Figure 3a) show the original AB stacked bilayer $MoS_2$ at $t = 0$ s. At $t = 7$ s, local atomic rearrangement and a relative shift between the two $MoS_2$ layers are observed (Figure 3b,f). The shift leads to the linear features as seen at the lower left corner of Figure 3b (enclosed in a pink box). The separation between these linear features (orange lines in Figure 3f) is $2.7 \pm 0.05$ Å, which is equivalent to the single-layer Mo lattice periodicity $a_1$ ($= \frac{\sqrt{3}}{2}a$, Figure 1). The simulated STEM ADF image (Supporting Information, Figure 3) shows that a relative in-plane displacement[1] $u = \frac{1}{3}a_1$ agrees well with the experiment. The lateral shift between layers can be a source of large local strain, which may trigger other structural rearrangement or defects. For example, for a local region, we observe obvious shrinkage of the projected Mo–Mo distance (green lines in Figure 3f). This lattice distortion is most likely caused by S vacancy line defects which have also been observed in single-layer $MoS_2$ samples under TEM[27,28] due to the "knock-on" effect by the electron beam, as the electron beam's energy (80 keV) is close to the threshold for S vacancy formation in single-layer $MoS_2$. Interstitial atoms and missing S-S columns are also observed in Figure 3f, as indicated by white arrows and blue arrows, respectively. These interstitial atoms and vacancies are highly mobile under combined electron irradiation and thermal heating, and eventually lead to the nucleation of AA′ stacked region shown in Figure 3g (blue hexagons). A straight stacking boundary (the orange line in Figure 3g) is formed between AA′ and AB stacked regions. We also notice that, once the AA′ stacked domain nucleates, the interlayer shift (which originally concentrates at the lower left corner in Figure 3b) significantly reduces. Therefore, it is likely that the nucleation and formation of the AA′ stacked region release the strain caused by the interlayer shift, and the decrease in elastic energy compensates for the energy gain due to the newly formed stacking boundaries.

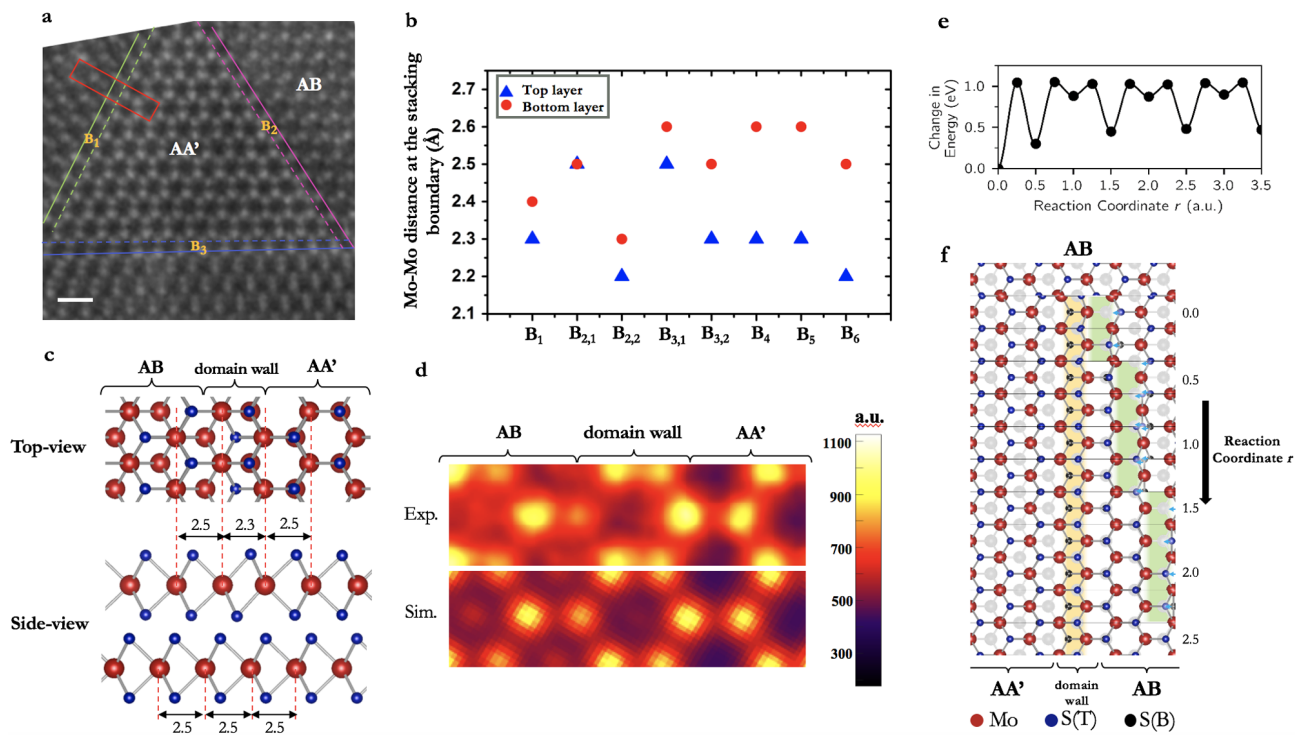

**Figure 4.** Atomic structure at the stacking boundaries. (a) A typical triangular AA′ stacked domain is connected to the AB stacked domain by atomically sharp boundaries. At the three boundaries (labeled as $B_1$, $B_2$, $B_3$), Mo lattices from the top layer are outlined by solid lines, and those from the bottom layer are outlined by dashed lines. Scale bar is 5 Å. (b) Summary of Mo lattice distances measured at six boundaries of two different samples in which AA′ and AB domains are stitched together (details of the boundary locations and labels can be found in the Supporting Information). Blue triangles represent the Mo lattice distances from the top layer, and red dots represent the Mo lattice distances from the bottom layer. The Mo lattice distances in the top layer at these boundaries shrink more than those in the bottom layer, indicating more severe structural change and thus more energy injection from the electron beam into the top layer. (c) Schematic of the boundary structure which is derived from DFT calculations and matches the experimentally observed most dominant boundary structure. This boundary structure features a T-phase-like structure. (d) Experimental STEM ADF image at a typical stacking boundary, with the intensity averaged over 5 unit cells across the boundary, compared to the simulated STEM ADF image based on the atomic model in (c). The experimental and simulated STEM ADF images are plotted on the same color scale. (e) Energy barrier per atom as S migrates during domain wall propagation. (f) Schematic diagram of domain wall nucleation ($r = 0.0$) and domain growth ($r > 0.0$), starting from a pristine, strained AB-stacked bilayer. See the Supporting Information Movie that depicts this process.

The AA′ stacked region grows under continuous electron beam irradiation, from ~2.88 nm² in Figure 3c to ~4.75 nm² in Figure 3d after 7 s. At this point, a second straight stacking boundary on the right (Figure 3h) forms at 60° with respect to the left boundary. Since the corner where these two boundaries meet is pinned by some defects, the AA′ stacked region cannot expand upward and has to grow downward in the direction indicated by vector $W$. It is worth noting that, when we image and heat the as-grown AA′ stacked bilayer $MoS_2$ under the same conditions, triangular AB stacked domains can also nucleate and grow from the originally AA′ stacked region (Supporting Information, Figure 4). This two-way transition between AA′ and AB stacked bilayer $MoS_2$ suggests that the difference in total energy between AA′ and AB stacked bilayer $MoS_2$ is very small, and excitations such as thermal energy and electron irradiation can overcome the energy barrier to allow the transition between these two stacking sequences. It is worth noting that we did not observe stacking boundary formation at a lower temperature such as 300 °C or a higher temperature (>400 °C, where Mo and S atoms can be knocked out by the electron beam easily). This indicates the thermal annealing at a certain temperature provides the right amount of energy to assist the motion of S atoms (which is the key step for the stacking sequence transition) while maintaining the crystal framework. In addition, we always observe stacking sequence transition seconds or tens of seconds after we start imaging an area, although we have let the sample stay at a certain temperature for a while (>30 min) before the imaging. This indicates the local interaction between the electron beam and the sample during the STEM imaging is the trigger for the stacking sequence transition.

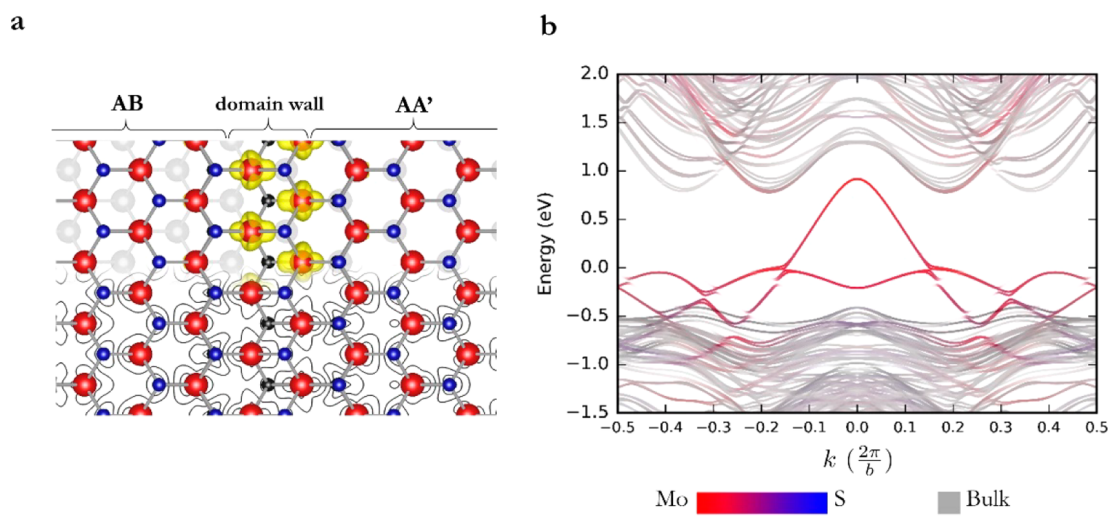

**Figure 5.** First-principles modeling of domain walls. (a) Atomic structure of the domain wall. The left side shows the AB domain, while the right side shows the AA′ domain, separated by the domain wall. The top half of the figure shows both the top and the bottom layers of the bilayer, superposed with the 15% isosurface of the charge densities of the boundary states. The bottom half of the figure shows only the top layer, superposed with the contour lines of the same charge densities on a two-dimensional cut of the Mo plane. (b) Band structure of the domain wall within DFT in the local density approximation (LDA).

We now examine the atomic structure of the stacking boundaries between the AB and AA′ stacked regions (Figure 4a). The atomic structure shown in the STEM images of these boundaries has three main characteristics: (1) Mo lattices are continuous from the AA′ stacked region to the AB stacked region, which means that the Mo atoms in both regions remain roughly at their original positions; (2) the hollow centers of the hexagonal lattices at the boundary are occupied by fewer S atoms compared to the original AB stacked bilayer $MoS_2$, as shown in the experimental STEM ADF image in Figure 4d; (3) the Mo−Mo distance near the boundary has decreased. These observations are possible because Mo-S-S columns and Mo columns in AB stacked bilayer $MoS_2$ show different intensities in ADF images. We then can differentiate the Mo atoms in the top layer from those in the bottom layer based on this intensity difference. In Figure 4a, three solid lines connect the Mo lattice points at the boundaries in the top layer (labeled as $B_1$, $B_2$, and $B_3$, respectively), while the dashed lines connect the Mo lattice points at the boundaries in the bottom layer. The Mo−Mo distances of both layers at these boundaries are shorter than that of bulk $MoS_2$, which has also been observed in monolayer $MoS_2$ with S line defects.[27,28] Averaged over six boundaries (see the Supporting Information, Figure 5 for the boundary locations), the average Mo−Mo distance at the stacking boundaries is ~2.3 Å in the top layer, and ~2.5 Å in the bottom layer (Figure 4b). This indicates the top layer and bottom layer may experience different energetics due to electron−matter interaction.

To further quantify and understand the exact atomic structure at the stacking boundaries, we perform first-principles DFT calculations[29] on different boundary structures (see the Supporting Information for computational details). The boundary structure in Figure 4c best satisfies the three experimentally observed characteristics mentioned above, and is stabilized by having a T-phase-like structure. The simulated STEM ADF image of this boundary structure agrees very well with the experimental results (Figure 4d). Moreover, our calculations show that compressive strain due to Mo−Mo distance shrinkage favors the formation of S vacancies (see the Supporting Information). This indicates that the nucleation of this T-phase-like boundary probably originates from the formation of S line defects, which are commonly observed in single-layer $MoS_2$ under extensive electron beam irradiation.[27,28] To understand each step of this boundary formation, we rely on DFT calculations to identify the minimum-energy path for the nucleation and growth of the AA′ stacked region from the original AB stacking (Figure 4f, only top $MoS_2$ layer is shown in color). Here, we use $r$ to represent the reaction coordinate of domain growth, and the energy change associated with each step is plotted in Figure 4e. At $r = 0.0$, a domain wall (shaded orange in Figure 4f) is formed by ejecting the top S atoms (blue balls) in the top $MoS_2$ layer. This leaves the bottom S atoms (black balls) exposed. Subsequent domain growth is achieved by the migration of the nearest S atoms in the top $MoS_2$ layer toward the domain wall. This process only involves S atoms in the top layer of bilayer $MoS_2$, because the energy absorbed by the bottom $MoS_2$ layer from the electron beam is significantly reduced due to the shielding effect by the top layer.[30] The minimum-energy path for the S migration in the top $MoS_2$ layer is composed of two steps: (1) the top S atoms of the top layer migrating first ($r = 0.25$), followed by (2) the bottom S atoms of the same layer migrating ($r = 0.5$), which is also a local energy minimum (Figure 4e). At this point, a unit cell of AA′ stacking is formed (Figure 4f). Subsequent expansion of the AA′ region involves the repetition of the above two S migration steps (Figure 4f). Since the energy barrier for each migration step is small, ranging from 0.15 to 1.05 eV per atom (Figure 4e), this domain growth process can be easily activated by the 80 keV electron beam used in the experiment. This S migrating process has also been previously reported for the H-to-H′ transition in monolayer $MoS_2$,[24] which is also observed in our experiment (Supporting Information, Figure 2a,b). Notably, this inversion domain formation in monolayer $MoS_2$ alone cannot explain our experimentally obtained atomically sharp stacking boundaries, which require both domain inversion in the top layer via S migrating and bonding modification in the bottom layer at the boundaries (Figure 4 and Supporting Information, Figure 2c). This indicates a strong interlayer interaction is the key to forming atomically sharp stacking boundaries in bilayer $MoS_2$. We emphasize the novel integration of different inversion symmetries in the same bilayer $MoS_2$ sample via creation of atomically sharp stacking boundaries between different stacking sequences.

Finally, we calculate the electronic band structure of the boundary and find in-gap states localized at the domain wall (see the Supporting Information for details). To identify states from the boundary, we project the band structure onto atomic wave functions of atoms near the stacking boundary (Figure 5a). In the band structure (Figure 5b), states with contributions from Mo and S atoms adjacent to the domain wall are colored red and blue, respectively, while contributions from atoms in the bulk are colored gray. We see that, importantly, the presence of the domain wall introduces metallic in-gap states that are absent in either AA′ or AB stacked $MoS_2$ bilayer. Similar in-gap states have also been observed at the edges[23,31] and mirror twin boundaries[20,23,32] of monolayer $MoS_2$. The in-gap states in our case are composed primarily of $4d$-states from Mo atoms bordering the domain wall. From the contour lines of the integrated charge density in Figure 5a, we see that these in-gap states are exponentially localized at the domain wall. We note that DFT-LDA has a well-known tendency to under-estimate band gaps and overestimate occupied band-widths.[33] Appropriate theory of higher levels such as the GW approximation[34−36] must be used to obtain quantitatively accurate quasiparticle band structures. Here, LDA already provides a good description of qualitative features and wave function character.

In summary, we have demonstrated a novel way to engineer domains with different inversion symmetries in bilayer $MoS_2$ through the synergy of thermal excitation and electron irradiation. Using our method, we are able to reversibly convert between regions of AA′ and AB stacking, thus switching between stacking sequences that result in different valley polarization. Through a first-principles study, we find that there are highly localized metallic states at the atomically sharp stacking boundaries. We propose that this approach may be used to reversibly pattern regions of different inversion symmetries, as well as controllably embed defect lines with different transport properties, within 2D semiconductors. This method of domain and symmetry engineering can be generalized to other TMDCs. We hope that this study will inspire greater interest in future studies of valleytronics control and defect state patterning in atomically thin $MoS_2$ and other TMDCs via nanoscale symmetry engineering.

## ■ ASSOCIATED CONTENT

**⊛ Supporting Information**

The Supporting Information is available free of charge on the ACS Publications website at DOI: 10.1021/acs.jpcc.7b08398.

> Experimental methods for CVD growth of $MoS_2$, TEM sample preparation, and STEM ADF imaging and multislice simulation. Identification of AB and AA′ stacked bilayer $MoS_2$ regions by comparing the experimentally collected and simulated STEM ADF images. Experimentally observed phase transition from H to H′ in monolayer $MoS_2$ and interlayer interaction in bilayer $MoS_2$. The formation of linear features in AB stacked bilayer $MoS_2$ due to the relative shift between the layers. Experimental observation of transition from AA′ stacking to AB stacking in bilayer $MoS_2$. Measurement of Mo−Mo distance at the stacking boundaries between AA′ and AB stacked domains. Details for first-principles calculations, including calculations of different stacking boundary structures between AA′ and AB stacked bilayer $MoS_2$ domains, band structures at these boundaries with and without H passivation, and the formation energy for S vacancy (PDF)
> Movie depicting domain wall nucleation and domain growth (MPG)
> Movie of continuous STEM ADF imaging showing the transition between different stacking sequences (MPG)

## ■ AUTHOR INFORMATION

**Corresponding Author**

*E-mail: azettl@berkeley.edu.

**ORCID** ⓘ

Aiming Yan: 0000-0002-5394-7498

Diana Y. Qiu: 0000-0003-3067-6987

**Author Contributions**

A.Z. and A.Y. conceived and designed the experiment. A.Y. collected and analyzed scanning transmission electron micros-copy (STEM) data. C.S.O., D.Y.Q., and S.G.L. performed density functional theory calculations. A.Y. and C.M. synthesized $MoS_2$ samples. C.O. wrote the Matlab code for multislice simulations. C.O. and J.C. contributed to and provided advice for STEM data collection and analysis. A.Y., A.Z., C.S.O., D.Y.Q., and S.G.L. co-wrote the manuscript. All authors discussed the results and commented on the paper.

**Notes**

The authors declare no competing financial interest.

## ■ ACKNOWLEDGMENTS

This work was supported in part by the Director, Office of Science, Office of Basic Energy Sciences, Materials Sciences and Engineering Division, of the U.S. Department of Energy under Contract No. DE-AC02-05-CH11231, within the sp2-bonded Materials Program (KC2207) which provided for STEM data collection and DFT calculations, and within the van der Waals Heterostructures Program (KCWF16) which provided for analysis of the STEM data; by the Molecular Foundry of the Lawrence Berkeley National Laboratory, supported by the Office of Science, Office of Basic Energy Sciences, of the U.S. Department of Energy under Contract No. DE-AC02-05CH11231 which provided for additional STEM instrumentation; by the National Science Foundation, under Grant No. 1542741, which provided for sample growth and preparation, under Grant No. DMR-1206512, which provided for development of the sample transfer method, and under Grant No. DMR-1508412, which provided for supplementary DFT calculations. Computational resources have been provided by the NSF through XSEDE resources at SDSC. C.S.O. acknowledges support from the Singapore National Research Foundation (Clean Energy) PhD Scholarship. We acknowledge Chengyu Song in the Molecular Foundry of the Lawrence Berkeley National Laboratory for TEM technical support.